# Attribution Locus and the Timeliness of Long-lived Asset Write-downs

Yao-Lin Chang

National Taipei University of Business

yaolinchang@ntub.edu.tw

Chun-Yang Lin (Corresponding Author)

National Taipei University of Business

Address: No.321, Sec. 1, Jinan Rd., Zhongzheng Dist., Taipei City 10051, Taiwan, ROC

Tel: 886-2-23226538

E-mail: chunyanglin@ntub.edu.tw

Fax: 886-2-23226395

Chi-Chun Liu

National Taiwan University

ccliu@ntu.edu.tw

Stephen G. Ryan

Stern School of Business, New York University

sryan@stern.nyu.edu

*This work was financially supported by the Center for Research in Econometric Theory and Applications (Grant no. 109L900202) from The Featured Areas Research Center Program within the framework of the Higher Education Sprout Project by the Ministry of Education (MOE) in Taiwan, and by the National Science and Technology Council of Taiwan
under Grant No. 102-2410-H-002-029-MY2, 104-2410-H-141-001, 109-2634-F-002-045, 112-2410-H-141-019.

*Data availability: Data used in this paper is publicly available.

# Attribution Locus and the Timeliness of Long-lived Asset Write-downs

**Abstract**: We examine the relative timeliness with which write-downs of long-lived assets incorporate adverse macroeconomic and industry outcomes versus adverse firm-specific outcomes. We posit that users of financial reports are more likely to attribute adverse firm-specific outcomes to suboptimal managerial actions, which provide managers with more incentive to delay write downs. We provide evidence that, controlling for other incentives to manage earnings, firms record write-downs in the current year that are driven by adverse macroeconomic and industry outcomes during both the current year and the next year, but they record write-downs driven by adverse firm-specific outcomes only in the current year.

1. **Introduction**

Declines in the values of assets below their carrying values may be driven by adverse outcomes that are macroeconomic, firm-specific, or of various intermediate levels of aggregation (e.g., industry-level). In this study, we posit that, relative to adverse macroeconomic and industry outcomes, users are more likely to attribute adverse firm-specific outcomes to suboptimal managerial actions, thereby providing managers with the incentive to delay asset write-downs driven by those outcomes. Consistent with this position, we hypothesize and provide evidence that asset write-downs incorporate adverse firm-specific outcomes on a less timely basis than adverse market-wide and industry outcomes, controlling for other incentives to manage earnings. Specifically, we find that firms record asset write-downs in the current year that are attributable to adverse macroeconomic and industry outcomes during both the current year and the next year, but they record write-downs attributable to adverse firm-specific outcomes only in the current year. That is, in determining asset write-downs, managers anticipate adverse macroeconomic and industry outcomes but not adverse firm-specific outcomes.

We empirically examine firms' reported write-downs of long-lived tangible and identifiable amortizing intangible assets (hereafter "asset write-downs") during 1996–2019, a period over which the accounting rules governing these write-downs were largely unchanged. These rules were contained in Statement of Financial Accounting Standards No. (FAS) 121, *Accounting for the Impairment of Long-Lived Assets and for Long-Lived Assets to Be Disposed Of*, from 1996 to 2001; in FAS 144, *Accounting for the Impairment or Disposal of Long-Lived Assets*, from 2002 to the adoption of the Accounting Standards Codification (ASC) in 2009; and in ASC 360-10-35 subsequently.

Our hypotheses are based on managers' differential incentive to delay asset write-downs driven by adverse macroeconomic and industry versus firm-specific outcomes. This variation stems from differences in the level of attribution locus for these types of adverse outcomes. We focus the empirical analysis on whether asset write-downs incorporate adverse macroeconomic and industry outcomes and adverse firm-specific outcomes with differential timeliness. We further conduct cross-sectional analyses that partition firms each year into above- and below-median groups based on proxies for attribution of adverse outcomes to managers. These analyses support the conclusion that greater attribution of adverse outcomes to managers induce more delay in asset write-downs driven by adverse firm-specific outcomes. Overall, our findings suggest that firms' recorded asset write-downs do not incorporate adverse firm-specific outcomes in a timely fashion either on a standalone basis or relative to the incorporation of adverse macroeconomic and industry outcomes.

Our findings contribute most directly to the literature on the determinants of asset write-downs, which we summarize in detail in Section 2.2. Much of this literature examines the extents to which write-downs are explained by economic variables versus managers' reporting incentives (e.g., Riedl, 2004). The approaches typically taken in this prior literature effectively treat managerial reporting incentives over the recording of asset write-downs as independent of the economic variables that give rise to the corresponding economic impairments. In contrast, we hypothesize and provide evidence that managerial reporting incentives strongly differ for economic impairments attributable to adverse macroeconomic and industry outcomes versus adverse firm-specific outcomes. Seond, largely motivated by economic theory, extensive prior literature provides evidence that managers make financial reporting decisions based on a diverse set of incentives, including litigation risk (e.g., Skinner, 1994, 1997), reputation preservation and development (e.g., Skinner, 1994), and career concerns (e.g., Baginski et al., 2018; Ali et al., 2019). Our findings suggest that psychological attribution theory, which posits that the strength of the incentives vary with the perceived level of managers' responsibility for outcomes, helps parsimoniously reconcile many of these diverse incentives. Our finding also contributes to the literature on the behavioral biases which affect financial reporting and disclosure decisions. Prior literature provides evidence that managers delay the disclosure of adverse outcomes when they are the likely attribution locus (Schloetzer et al., 2020). In contrast, our study focus on the biases in financial statement recognition. When prior literature provides evidence that managers exhibit less strategic bias under recognition than disclosure (Clor‐Proell and Maines, 2014), our finding suggests that phychological attribution locus still affects managers' asset write-down reporting and has implications in light of the research on managers' reporting behavior.

## 2. Literature review and hypotheses

### 2.1. Impact of attribution locus on managers' reporting behavior

Attribution theory in psychology models how observers (in our setting, users of financial reports) attribute causal explanations for events. Heider (1958) proposes a simple dichotomy of the loci for such attributions: internal versus external. This dichotomy bears directly on whether the individuals involved should be blamed/punished for adverse outcomes or lauded/rewarded for favorable outcomes. External attribution reduces or eliminates the individuals' responsibility for the outcomes (Kelley and Michela, 1980; Aerts, 2005; Koonce et al., 2011).

Prior literature provides evidence that firm managers engage in financial reporting behavior consistent with attribution theory. For example, relative to their attributions of favorable outcomes, managers are more likely to attribute the cause of adverse outcomes to external circumstances (Baginski et al., 2000; Baginski et al., 2004). Our research question is most closely related to Schloetzer et al. (2020) that provide evidence that firms are less likely to disclose information regarding a negative economic event for which the firm is likely to be blamed than a negative event for which the firm is likely to be perceived as blameless. First, as in our study, Schloetzer et al. (2020) examine the disclosure propensity following a contemporaneous negative economic event, but unlike our study they do not focus on the disclosure propensity for future outcomes. Second, Schloetzer et al. (2020) examine the difference in disclosure propensity between blamed and blameless events. However, our study examine the difference in disclosure propensity between blamed and blameless outcomes within the same financial reporting. Third, Schloetzer et al. (2020) examine the disclosure following a negative event but our study focus on the financial statement recognition of adverse economic outcomes.

## 2.2. Accounting for write-downs of long-lived tangible and identifiable amortizing intangible assets

Extensive prior literature provides evidence that managers act in their self-interest based on various incentives in determining whether to record asset write-downs in a period. For example, newly installed managements often record asset write-downs to "clear the decks" or to attribute the write-downs to their predecessors, and they overstate these write-downs to create "reserves" that can be released in future periods (e.g., Strong and Meyer, 1987; Elliott and Shaw, 1988; Francis et al., 1996; Beatty and Weber, 2006; Ramanna and Watts, 2012). Asset write-downs are also involved in "big baths" in bad times or in smoothing earnings downward in good times (e.g., Zucca and Campbell, 1992). Somewhat in contrast, other studies provide evidence that firms record asset write-downs when circumstances indicate that they should. Distressed firms are more likely to record asset write-downs (e.g., Elliott and Shaw, 1988; Elliott and Hanna, 1996), asset write-downs are associated with future negative abnormal accruals (e.g., Rees et al., 1996), and changes in sales and operating cash flows explain tangible asset write-downs (e.g., Banker et al., 2017).

Our research question and design are most closely related to two prior studies that provide evidence that the GAAP governing asset write-downs described above does not work as intended due to the non-recoverability trigger for impairment. First, as in our study, Riedl (2004) examines the associations of asset write-downs with macroeconomic, industry, and firm-specific outcomes, but unlike our study he does not focus on differences in these associations or the reasons for these differences. Second, Gordon and Hsu (2018) investigate the differential ability of write-downs of long-lived tangible assets under IFRS versus GAAP to predict future operating cash flows. However, Gordon and Hsu (2018) do not link the lack of predictive power of asset write-downs under GAAP to firms delaying the recording of asset write-downs driven by adverse firm-specific outcomes, as our findings suggest.

## 2.3. Hypotheses

Consistent with psychological attribution theory discussed in Section 2.1 as well as with prior accounting research discussed below, we posit that market participants are more likely to hold managers responsible for adverse firm-specific outcomes than for adverse macroeconomic and industry outcomes (e.g., Verrecchia, 1983, 1986; Watts and Zimmerman, 1986; Bratten et al., 2016). Because adverse outcomes for which managers are held responsible are likely to adversely affect their careers (e.g., reduce their compensation, retention and promotion prospects, and outside opportunities) and overall reputations, we further posit that managers have stronger incentives to delay the reporting of such outcomes, either because their horizons are sufficiently short or because they gamble that conditions will improve (e.g., Verrecchia, 2001; Kothari et al., 2009; Baginski et al., 2018; Ali et al., 2019).

Based on the above, we expect that managers have substantial incentive to delay asset write-downs driven by adverse firm-specific outcomes, but little incentive to delay write-downs driven by adverse macroeconomic and industry outcomes. Specifically, while accounting rules require managers to anticipate all future adverse outcomes in determining the write downs of assets for which the carrying values are deemed uncollectible, we expect managers to anticipate future adverse macroeconomic and industry outcomes more than future adverse firm-specific outcomes in determining these write-downs. We state these expectations formally in the following hypotheses.[1]

> **H1M:** Asset write-downs incorporate current and future adverse macroeconomic and industry outcomes.
>
> **H1F-S:** Asset write-downs incorporate current adverse firm-specific outcomes but not future adverse firm-specific outcomes.

## 3. Models and variables

H1M and H1F-S examine whether asset write-downs in the current year incorporate adverse macroeconomic, industry, and firm-specific outcomes, respectively, in the next year. We test hypotheses H1M and H1F-S using a model with the following structure. The dependent variable is either an indicator for firm $i$ recording asset write-down during year $t$ ($WO_{i,t}$) or the amount of that asset write-down (expressed as a positive number, $WOTA_{i,t}$). See Appendix A for detailed definitions of all model variables.

---

[1] Our main hypothesis rests on the premise that investors will treat the same amount of asset write-downs differently when write-downs are due to different attribution locus. Prior literature provides evidence that analysts and investors respond to managers' financial reporting consistent with attibution thoery (Barton and Mercer 2005; Kimbrough and Wang 2014). More distantly, prior literature also provides evidence that changes in managerial credibility which are measured by the market reponses and analysts' reponses to earnings announcements after their withdrawal disclosure differ based on the attribution provided by managerment (Marshall and Skinner 2022).

Following Riedl (2004) and Klein and Marquardt (2006), our proxy for the macroeconomic outcome in year $t$ is the percentage change in U.S. GDP in that calendar year ($\Delta GDP_t$). Our proxy for the industry outcome in year t is the median change in industry percentage change in sales in that calendar year ($\Delta INDSALE_t$). We expect negative values of $\Delta GDP_t$ and $\Delta INDSALE_t$ to be associated with declines in asset values that might drive asset write-downs.

To capture multiple key dimensions of firm performance, our proxy for the firm-specific outcome for firm $i$ in year $t$ is the first principal component of the changes in the firm's sales, earnings before asset write-downs and depreciation expenses, and operating cash flows during year $t$ ($FIRM_{i,t}$).[2] We expect negative values of $FIRM_{i,t}$ to be associated with declines in asset values that may drive asset write-downs.

Drawing on extensive prior literature (e.g., Strong and Meyer, 1987; Elliott and Shaw, 1988; Francis et al., 1996; Riedl, 2004; Beatty and Weber, 2006; Ramanna and Watts, 2012; Zucca and Campbell, 1992), we control for four managerial reporting incentive variables: management changes ($\Delta MGT_{i,t}$), earnings-smoothing ($SMOOTH_{i,t}$); big-bath ($BATH_{i,t}$), and avoiding violations of debt covenants ($DEBT_{i,t}$).

Based on the above, we test H1M and H1F-S using the following model:

---

[2] We use pre-write down and depreciation earnings as the proxy for firm performance to alleviate the concerns for reverse causality issue.

$$WOTA_{i,t} \text{ or } WO_{i,t} = \alpha_0 + \beta_1 \Delta GDP_t + \beta_2 \Delta GDP_{t+1} + \beta_3 \Delta INDSALE_{i,t} + \beta_4 \Delta INDSALE_{i,t+1} + \beta_5 FIRM_{i,t} + \beta_6 FIRM_{i,t+1} \quad (1)$$
$$+ \beta_7 \Delta MGT_{i,t} + \beta_8 SMOOTH_{i,t} + \beta_9 BATH_{i,t} + \beta_{10} DEBT_{i,t} + \varepsilon_{i,t}.$$

We estimate equation (1) with dependent variable $WOTA_{i,t}$ ($WO_{i,t}$) using tobit (logit). Based on H1M, we expect $\beta_1, \beta_2, \beta_3$ and $\beta_4$ to be negative. Based on H1F-S, we expect $\beta_5$ to be negative and $\beta_6$ to be zero.[3,4]

## 4. Sample and descriptive statistics

We obtain accounting data from Compustat's Fundamental Annual Database (Xpressfeed format), changes in firms' top three managers from Execucomp, and GDP from the website of the Bureau of Economic Analysis. The sample includes all firms except for financial firms (two-digit Global Industry Classifications Standard ["GICS"] sector code 40). Because FAS 121 became effective in 1996, our asset write-down data begins in 1996. Our data for all variables end in 2019.

Our sample period begins in 1997 to exclude any possible impacts of SFAS 121 transition period and ends in 2018 because our models include outcomes in year *t+1* (thus losing 2019). Observations with missing data on any model variable other than asset write-downs are excluded from the sample, yielding a final sample of 30,111 firm-year observations. Continuous variables are winsorized at the bottom and top one percent of their distributions. Table 1 shows the summary statistics of all the variables used in the current study.

---

[3] The macroeconomic and industry variables included in our model also alleviate potential confounding effect due to omitted common shocks which affect both asset write-downs and firm specific outcomes (Whited, Swanquist, Shipman, and Moon 2022).

[4] Prior empircal accounting research investigates the predictive value of accounting numbers or treatments for future performance of firms with drawing causal links between the accounting numbers or treatments and one-year-ahead performance measures (e.g., Fairfield, Sweeney, and Yohn 1996, Barth et al. 2001; Gordon and Hsu 2018). Our model specification which examine the t and t+l relation for macro, industry, and firm is consistent with the approach taken in the recent literature.

# 5. Empirical results

## 5.1. Main Results

Table 2 presents the tests of H1M and H1F-S based on the estimations of equation (1) with $WOTA_{i,t}$ and $WO_{i,t}$ as the dependent variable in the left and right columns, respectively. The coefficients on $\Delta GDP_t$, $\Delta GDP_{t+1}$, $\Delta INDSALE_t$ and $\Delta INDSALE_{t+1}$ are significantly negative at the five percent level or better in both columns, consistent with H1M that asset write-downs incorporate expectations about both current and future adverse macroeconomic and industry outcomes. The coefficient on $FIRM_{i,t}$ is significant at the one percent level and the coefficient on $FIRM_{i,t+1}$ is insignificant in both columns, consistent with H1F-S that asset write-downs incorporate current but not future adverse firm-specific outcomes.[5]

## 5.2. Additional Tests

In this section, we conduct cross-sectional tests to identify the impacts of attribution locus on the estimation of equation (1) reported in Table 2. We construct the proxy for attribution locus in two ways. First, prior research shows that firms often explicitly evaluate and compensate managers based on their performance relative to specified peer groups (e.g., Aggarwal and Samwick, 1999; Blackwell et al., 1994; DeFond and Park, 1999), and that managers that perform poorly relative to these groups tend to engage in opportunistic behavior (e.g., Lewellen et al., 1996; Dikolli et al., 2018; Gong et al., 2019). Based on this prior research, we expect that managers are more likely to be viewed as the attribution locus for asset write-downs when their firms' performance falls below that of their industry peers. We use the first principal component of three industry-adjusted performance variables—change in sales ($\Delta SALE_{i,t}$), change in earnings before asset write-downs and depreciation expense ($\Delta E_{i,t}$), and change in cash flows from operations ($\Delta OCF_{i,t}$)—to capture the likelihood that users of financial reports attribute adverse outcomes to firm managers. We define industries as two-digit GICS sectors, and denote this first principal component by $ATTRIB_{i,t}$.

---

[5] Our results remain robust to the use of alternative proxies and the inclusion of additional control variable. First, we replace the proxy for macroeconomic outcomes in equations (1), $\Delta GDP_t$, with the first principal component of four macroeconomic variables: $\Delta GDP_t$; the percentage change in the employment rate ($\Delta EMPLOY_t$), the change in total industrial production ($\Delta IP_t$), and (4) the change in the consumer price index ($\Delta CPI_t$). Second, we replace the change in sales with the change in cash sales to ensure that our results are not driven by the accrual component of sales. We also replace the change in operating cash flows with the level of operating cash flows. Third, we include the begnning-of-period market-to-book ratio to control buffers to write-downs.

Second, we expect users to be less likely to attribute asset write-downs to newly installed firm managements, because these managements can (and often do) blame prior managements for the write-downs (e.g., Strong and Meyer, 1987). We capture this effect using an indicator for whether firms have experienced management changes during the year, denoted $\Delta MGT_{i,t}$. $D\_ATTRIB_{i,t}$ indicates either an above-median value of $ATTRIB_{i,t}$ or $\Delta MGT_{i,t}$ taking a value of one, that is, low attribution of adverse outcomes to the firm's managers. To test whether managers are less likely to be viewed as the attribution locus for asset write-downs record more timely asset write-downs driven by firm-specific outcomes, we expand equation (1) by including interactions of $D\_ATTRIB_{i,t}$ and the future firm-specific outcome variables.[6]

Results are reported in Table 3. As expected, the coefficient on $FIRM_{i,t+1} \times D\_AATRIB_{i,t}$ is significantly negative at the ten percent level or better in all four columns, consistent with asset write-downs incorporating future adverse firm-specific outcomes when users of financial reports are less likely to attribute to the write-downs to managers.

---

[6] We do not include interactions of $D\_ATTRIB_{i,t}$ and adverse current macroeconomic, industry, and firm-specific outcomes because these outcomes are readily observable and subject to less information asymmetry between firm managers and users of financial reports, thereby providing managers less ability to delay asset write-downs driven by these outcomes. We do not include interactions of $D\_ATTRIB_{i,t}$ and adverse future macroeconomic and industry outcomes because users are less likely to attribute these outcomes to suboptimal managerial actions, thereby providing managers with less incentive to delay asset write-downs driven by those outcomes (Baginski et al., 2000). When we include interactions of $D\_ATTRIB_{i,t}$ and adverse current and future macroeconomic, industry, and firm-specific outcomes, the estimated coefficients on the interaction term, $FIRM_{i,t+1} \times D\_ATTRIB_{i,t}$, remain negative and significant in all models.

## 6. Conclusion

In this study, we empirically examine firms' reported write-downs of long-lived tangible and identifiable amortizing intangible assets during 1996–2019 under FAS 121, FAS 144, and ASC 360-10-35. These accounting rules include a loose recoverability trigger for asset write-downs that provides managers with discretion to delay write-downs. We provide new insights as to when managers are incentivized to exercise such discretion. Specifically, we posit that, relative to adverse macroeconomic and industry outcomes, adverse firm-specific outcomes are more likely to be attributed by users of financial reports to suboptimal managerial actions. Hence managers have greater incentive to delay asset write-downs driven by firm-specific outcomes. Consistent with this position, we hypothesize and provide evidence that asset write-downs incorporate adverse firm-specific outcomes on a less timely basis than adverse market-wide and industry outcomes, controlling for other incentives to manage earnings.

Our findings contribute to findings by Riedl (2004) and Gordon and Hsu (2018) suggesting that the recoverability trigger for impairments of tangible and identifiable amortizing intangible assets in FAS 121, FAS 144, and ASC 360-10-35 deteriorates the timeliness of write-downs of these assets. As discussed in the introduction, our findings also contribute to the literature on behavior bias in financial reporting.

# Appendix A
# Variable Definitions

| | |
|---|---|
| *ATTRIB* | The first principal component of three firm-specific variables that indicate attribution of adverse outcomes to the managers of firm $i$ in year $t$: industry-adjusted changes in sale, industry-adjusted changes in pre-write downs and depreciation expense earnings, and industry-adjusted changes in operating cash flows. |
| $BATH_{i,t}$ | An indicator variable that equals 1 if the percentage change in firm $i$'s pre-write-down earnings from year $t-1$ to $t$ is below the median of the negative values of this variable over the same period, and 0 otherwise |
| $D\_ATTRIB_{i,t}$ | An indicator variable that equals 1 if *ATTRIB* is above-median of the variable across all firms in the industry or $\Delta MGT$ taking a value of one in year $t$ |
| $DEBT_{i,t}$ | The change in long-term debt for firm $i$ from year $t-1$ to $t$, divided by total assets at the end of year $t-1$ |
| $\Delta E_{i,t\ (i,t+1)}$ | The change in pre-write down and depreciation earnings for firm $i$ from year $t-1$ to $t$ ($t$ to $t+1$) divided by total common equity at the end of year $t-1$ ($t$) |
| $FIRM_{i,t\ (i,t+1)}$ | The first principal component of three firm-specific performance variables for firm $i$ in year $t$ ($t+1$): $\Delta SALE_{i,t\ (i,t+1)}$, $\Delta E_{i,t\ (i,t+1)}$, and $\Delta OCF_{i,t\ (i,t+1)}$ |
| $\Delta GDP_{t\ (t+1)}$ | The percentage change in U.S. gross domestic product from year $t-1$ to $t$ ($t$ to $t+1$) |
| $\Delta INDSALE_{t\ (t+1)}$ | The median change in firm $i$'s industry percentage change in sales for from year $t-1$ to $t$ ($t$ to $t+1$) in the industry-year |
| $\Delta MGT_{i,t}$ | An indicator variable that equals 1 if at least one of the top three compensated positions in firm $i$ changes in year $t$, and 0 otherwise |
| $\Delta OCF_{i,t\ (i,t+1)}$ | The change in cash flows from operations for firm $i$ from year $t-1$ to $t$ ($t$ to $t+1$) divided by total common equity at the end of year $t-1$ ($t$) |
| $\Delta SALE_{i,t\ (i,t+1)}$ | The change in sales for firm $i$ from year $t-1$ to $t$ ($t$ to $t+1$) divided by total common equity at the end of year $t-1$ ($t$) |
| $SMOOTH_{i,t}$ | An indicator variable that equals 1 if the percentage change in firm $i$'s pre-write-down earnings from year $t-1$ to $t$ is above the median of the positive values of this variable over the same period, and 0 otherwise |
| $WO_{i,t-1\ (i,t)}$ | An indicator variable that equals 1 if firm $i$ records a write-down of long-lived tangible and identifiable amortizing intangible assets in year $t-1$ ($t$), and 0 otherwise |
| $WOTA_{i,t}$ | Net-of-tax write-downs of long-lived tangible and identifiable amortizing intangible assets by firm $i$ in year $t$ (expressed as a positive number), divided by total assets at the end of year $t-1$ |

# Table 1

## Descriptive Statistics

## Full Sample and Subsamples with and without Asset Write-downs in Year t

**Panel A: Continuous variables**

| | Full sample (n = 30,111) | | | Asset write-down in year t (n = 4,567) | | | No asset write-down in year t (n = 25,544) | | | Differences across asset write-down and no write-down subsamples | |
|---|---|---|---|---|---|---|---|---|---|---|---|
| | Mean | Median | Std | Mean | Median | Std | Mean | Median | Std | Mean | Median |
| $WOTA_{i,t}$ | 0.001 | 0.000 | 0.004 | 0.007 | 0.003 | 0.009 | 0.000 | 0.000 | 0.000 | 0.007*** | 0.003### |
| $\Delta GDP_t$ | 0.024 | 0.025 | 0.016 | 0.019 | 0.022 | 0.015 | 0.025 | 0.026 | 0.016 | −0.006*** | −0.004### |
| $\Delta INDSALE_t$ | 0.136 | 0.144 | 0.133 | 0.100 | 0.117 | 0.132 | 0.142 | 0.150 | 0.132 | −0.042*** | −0.033### |
| $FIRM_{i,t}$ | −0.009 | −0.127 | 0.885 | −0.150 | −0.212 | 0.912 | 0.016 | −0.112 | 0.877 | −0.166*** | −0.100### |
| $\Delta DEBT_{i,t}$ | 0.027 | 0.000 | 0.121 | 0.018 | 0.000 | 0.111 | 0.029 | 0.000 | 0.122 | −0.011*** | 0.000### |

**Panel B: Indicator variables**

| | Full sample | | Asset write-down in year t | | No asset write-down in year t | | Differences across asset write-down and no write-down subsamples |
|---|---|---|---|---|---|---|---|
| | (n = 30,111) | | (n = 4,567) | | (n = 25,544) | | |
| Value of indicator = | 1 | 0 | 1 | 0 | 1 | 0 | |
| $\Delta MGT_{i,t}$ | 18,687 (62.06%) | 11,424 | 3,019 (66.10%) | 1,548 | 15,668 (61.34%) | 9,876 | 0.048 *** |
| $SMOOTH_{i,t}$ | 8,095 (26.88%) | 22,016 | 1,180 (25.84%) | 3,387 | 6,915 (27.07%) | 18,629 | -0.012 * |
| $BATH_{i,t}$ | 4,755 (15.79%) | 25,356 | 1,133 (24.81%) | 3,434 | 3,622 (14.18%) | 21,922 | 0.106 *** |

\*\*\*, \*\*, and \* (###, ##, and #) denote significance at 1%, 5%, and 10% levels, respectively, in two-tailed t-tests (Wilcoxon tests) of differences in means (medians).

**Table 2**
**Tests of H1M and H1F-S**
**Do Asset Write-downs Differentially Incorporate Adverse Macroeconomic versus Firm-specific Outcomes?**

|  | Predicted sign | | Dependent variable | |
|---|---|---|---|---|
|  | (Hypothesis) | | $WOTA_{i,t}$ | $WO_{i,t}$ |
| $\Delta GDP_t$ | − | (H1M) | -0.075*** | -9.884*** |
|  |  |  | (0.000) | (0.000) |
| $\Delta GDP_{t+1}$ | − | (H1M) | -0.056*** | -8.230*** |
|  |  |  | (0.000) | (0.000) |
| $\Delta INDSALE_t$ | − | (H1M) | -0.010*** | -1.072*** |
|  |  |  | (0.000) | (0.000) |
| $\Delta INDSALE_{t+1}$ | − | (H1M) | -0.003** | -0.342** |
|  |  |  | (0.014) | (0.044) |
| $FIRM_{i,t}$ | − | (H1F-S) | -0.001*** | -0.088*** |
|  |  |  | (0.000) | (0.000) |
| $FIRM_{i,t+1}$ | 0 | (H1F-S) | -0.000 | -0.024 |
|  |  |  | (0.593) | (0.215) |
| $\Delta MGT_{i,t}$ |  |  | 0.002*** | 0.187*** |
|  |  |  | (0.000) | (0.000) |
| $\Delta DEBT_{i,t}$ |  |  | -0.004*** | -0.592*** |
|  |  |  | (0.000) | (0.000) |
| $SMOOTH_{i,t}$ |  |  | 0.002*** | 0.200*** |
|  |  |  | (0.000) | (0.000) |
| $BATH_{i,t}$ |  |  | 0.006*** | 0.597*** |
|  |  |  | (0.000) | (0.000) |
| Constant |  |  | -0.013*** | -1.446*** |
|  |  |  | (0.000) | (0.000) |
| Chi-square test |  |  | 1,124.65*** | 923.25*** |
| Observations |  |  | 30,111 | 30,111 |

This table reports the tobit (logit) estimation of the $WOTA_{i,t}$ ($WO_{i,t}$) version of equation (1). Appendix A defines all model variables. ***, **, and * denote significance at the 1%, 5%, and 10% levels, respectively. Coefficient *p*-values are also presented in parentheses.

**Table 3**

**Impact of Lesser Attribution to Managers**

| | $D\_ATTRIB_{i,t}$ = | $ATTRIB_{i,t}$ | | $\Delta MGT_{i,t}$ | |
|---|---|---|---|---|---|
| | Dep. var. = | $WOTA_{i,t}$ | $WO_{i,t}$ | $WOTA_{i,t}$ | $WO_{i,t}$ |
| | Predicted sign | (1) | (2) | (5) | (6) |
| $D\_ATTRIB_{i,t}$ | | -0.001*** | -0.137*** | 0.002*** | 0.185*** |
| | | (0.000) | (0.000) | (0.000) | (0.000) |
| $\Delta GDP_t$ | | -0.068*** | -8.956*** | -0.074*** | -9.672*** |
| | | (0.000) | (0.000) | (0.000) | (0.000) |
| $\Delta GDP_{t+1}$ | | -0.059*** | -8.913*** | -0.057*** | -8.750*** |
| | | (0.000) | (0.000) | (0.000) | (0.000) |
| $\Delta INDSALE_t$ | | -0.010*** | -1.190*** | -0.012*** | -1.366*** |
| | | (0.000) | (0.000) | (0.000) | (0.000) |
| $\Delta INDSALE_{t+1}$ | | -0.004*** | -0.310# | -0.004** | -0.270# |
| | | (0.006) | (0.061) | (0.015) | (0.093) |
| $FIRM_{i,t}$ | | -0.000*** | -0.054** | -0.001*** | -0.090*** |
| | | (0.007) | (0.019) | (0.000) | (0.000) |
| $FIRM_{i,t+1}$ | | 0.000 | -0.003 | 0.000 | 0.026 |
| | | (0.471) | (0.920) | (0.219) | (0.440) |
| $FIRM_{i,t+1} \times D\_ATTRIB_{i,t}$ | − | -0.000* | -0.050# | -0.001* | -0.074* |
| | | (0.087) | (0.094) | (0.057) | (0.066) |
| Other controls | | Yes | Yes | Yes | Yes |
| Constant | | -0.012*** | -1.369*** | -0.013*** | -1.445*** |
| | | (0.000) | (0.000) | (0.000) | (0.000) |
| Chi-square test | | 1,150.79*** | 937.58*** | 1,131.15*** | 926.33*** |
| Observations | | 30,111 | 30,111 | 30,111 | 30,111 |

This table reports the tobit (logit) estimation of the $WOTA_{i,t}$ ($WO_{i,t}$) version of equation (4). $D\_ATTRIB_{i,t}$ indicates of low attribution of adverse outcomes to managers, either above-median $ATTRIB_{i,t}$ or $\Delta MGT_{i,t}$ equal to one. Appendix A defines all model variables. ***, **, and * denote significance at the 1%, 5%, and 10% levels, respectively, in two-tailed tests. # denote significance at the 10% levels in one-tailed tests. Coefficient $p$-values are also presented in parentheses.